\documentclass[aps,prb,amsmath]{revtex4}
\usepackage{graphicx}% Include figure files
\usepackage{bm}% bold math
\usepackage{bbold}

\def\br{ \bm{r} }

\def\im{ \mathrm{Im}\, }

\begin{document}
\title{Current-carrying states in FFLO superconductors}

\author{K. V. Samokhin and B. P. Truong}

\affiliation{Department of Physics, Brock University, St. Catharines, Ontario L2S 3A1, Canada}
\date{\today}

\begin{abstract}
We show that nonuniform superconductors of the Fulde-Ferrell-Larkin-Ovchinnikov (FFLO) type conduct electric current in the way which is very different from the usual case. We discuss both equilibrium and nonequilibrium 
properties using a modified Ginzburg-Landau formalism. Among the novel features are the existence of two different critical currents and two distinct stable states able to carry a given current,  
the possibility of superconducting domain walls, and also a spontaneous supercurrent in a ring geometry.   
\end{abstract}

\maketitle

\section{Introduction}
\label{sec: Intro}

It has been known since the seminal works by Fulde and Ferrell\cite{FF64} (FF) and Larkin and Ovchinnikov\cite{LO64} (LO) that the competition between the pair condensation energy and the ``paramagnetic'' pair 
breaking due to the spin splitting of the electron bands can result in the formation of peculiar nonuniform superconducting states, known as the FFLO (or LOFF) states.  
In contrast to the standard Bardeen-Cooper-Schrieffer (BCS) model, the Cooper pairs in an FFLO superconductor have a nonzero center-of-mass momentum, resulting in a spatial modulation of the order parameter. 
While in the simplest case this modulation is described by a single plane wave, $\psi(\br)\propto e^{i\bm{q}\br}$, which is known as the FF state, more complicated structures, 
containing two or more plane waves, such as the LO state with $\psi(\br)\propto\cos\bm{q}\br$, are also possible. 
Over the years, a number of superconducting materials have been suggested as possible hosts of the FFLO state, see Ref. \onlinecite{FFLO-review-07} for a review, however a definitive experimental confirmation is still lacking. 
More recently, alternative routes to the FFLO state have been discussed in the broader context of fermionic systems with a pairing instability and mismatched Fermi surfaces, 
such as ``cold'' Fermi gases\cite{RS09} and color superconducting quark matter.\cite{CasNar04} Most of the theoretical studies of the FFLO state focused on the mean-field phase diagram and 
the order parameter structure, but a number of works have also looked at the effects of fluctuations, see Refs. \onlinecite{Shima98,SM06,KCB07,Sam11}.

The goal of the present work is twofold. First, we investigate how an FFLO superconductor actually conducts
a constant electric current. Suprisingly, this simple question seems to have been largely overlooked in the literature. We would like to mention Ref. \onlinecite{DSK06}, where a current-driven FFLO instability was found 
in a superconductor with a nested Fermi surface, and also Ref. \onlinecite{BMS07}, where the critical current in an LO-type state was calculated. We focus on the quasi-one-dimensional (1D) case, which can be realized in a wire whose
transverse dimension is less than both the superconducting coherence length and the magnetic field penetration depth. The electron spectrum in this system is assumed to be three-dimensional, but the order parameter depends only on 
the coordinate $x$ along the wire. Our second goal is to develop a theoretical framework for studying the FFLO superconductors out of equilibrium and apply it to some select problems. While the nonequilibrium physics of 
the usual BCS superconductors is rich and complicated, see Refs. \onlinecite{IK84} and \onlinecite{Kopnin-book}, the analogous issues in the FFLO case have remained essentially unexplored.  

Properties of the FFLO superconductors, both equilibrium and nonequilibrium, can be studied at the mean-field level using a modification of the phenomenological 
Ginzburg-Landau (GL) formalism. In contrast to the usual case, the coefficient in front of the $|\nabla_x\psi|^2$ term in the GL free energy functional is negative, producing the superconducting instability 
with a nonzero wavevector. Microscopic derivation in the simplest model of 
a clean paramagnetically-limited isotropic superconductor shows that the gradient term indeed changes sign in a sufficiently strong magnetic field, but it does so simultaneously with the coefficient in front 
of the $|\psi|^4$ term. Therefore, in order to ensure stability, one has to include higher-order terms, such as $|\nabla_x^2\psi|^2$, $|\psi|^6$, and others (Ref. \onlinecite{BK97}).
Then the most stable state in 1D corresponds to a nonlinear generalization of the LO state, with the gap magnitude periodically modulated in space. On the other hand, it has been shown that if one takes into account 
disorder and the pairing anisotropy, then the $|\psi|^4$ term can remain positive while the $|\nabla_x\psi|^2$ term changes sign, thus making the FF state energetically more favorable than the LO state.\cite{AY01,HM06}
Motivated by this possibility as well as by the analytical simplicity of the calculations, in this work we focus on the properties of the FF superconducting state. 

The rest of the paper is organized as follows. In Sec. \ref{sec: equilibrium}, we study the current-carrying FF states in equilibrium, discussing, in particular, the critical currents, the stability issues, the topological defects 
(superconducting domain walls), and also the FF states in a ring geometry. In Sec. \ref{sec: TDGL}, we look into some nonequilibrium properties of the current-carrying states, using a modified time-dependent GL formalism.  
Sec. \ref{sec: Conclusions} concludes with a summary of our results. Throughout the paper $e$ denotes the absolute value of the electron charge.

\section{Equilibrium properties}
\label{sec: equilibrium}

Formation of a nonuniform FFLO state in a thin superconducting wire of length $L$ can be described by a modified GL functional ${\cal F}=\int_0^L F dx$, with the Helmholtz free energy density given by
\begin{equation}
	F = \alpha |\psi|^2 + \frac{\beta}{2}|\psi|^4 + K|\nabla_x\psi|^2 + \tilde{K}|\nabla_x^2 \psi|^2,
    \label{modified-GL-energy-density}
\end{equation}
where $\alpha=a(T-T_{c,0})$, $a>0$, and $T_{c,0}$ is the critical temperature of the transition into a uniform superconducting state. In the 1D case, the order parameter depends only on $x$
and the orbital effects of magnetic field can be neglected. In the analysis below, we neglect the boundaries and effectively consider 
an infinitely long wire (with the exception of Sec. \ref{sec: FF-ring}). In order for the instability with a finite wavevector to occur, one has to assume $K<0$, while
the last term, with $\tilde K>0$, is needed to stabilize the order parameter modulation.

The critical temperature of a nonuniform superconducting state is found by solving the linearized GL equation 
\begin{equation}
\label{linear-GL}
  \alpha \psi - K\nabla_x^2 \psi + \tilde{K}\nabla_x^4 \psi = 0.
\end{equation}
The particular solution is a plane wave $\psi(x)\propto e^{iqx}$, and the wavevector corresponding to the maximum critical temperature, $T_c=T_{c,0}+K^2/4a\tilde K$, is given by
\begin{equation}
\label{FFLO-wavevector}
  |q|=q_0=\sqrt{\frac{|K|}{2\tilde K}}.
\end{equation}
The actual state that is realized below $T_c$ is determined by the higher-order terms in the GL energy functional. As mentioned in the Introduction, with $\beta>0$ the single-plane-wave state (the FF state) 
\begin{equation}
\label{FF-state}
  \psi(x)\propto e^{\pm iq_0x}
\end{equation}
has lower energy than the LO state $\psi(x)\propto\cos(q_0x)$, see Appendix \ref{app: FF vs LO}. 

The supercurrent can be obtained from Eq. (\ref{modified-GL-energy-density}) in the standard fashion, by introducing the gauge-covariant derivative $\nabla_x\to\nabla_x+i(2e/\hbar c)A_x$, varying the
free energy with respect to the vector potential $A_x$, and subsequently setting $A_x=0$, with the following result:
\begin{equation}
\label{supercurrent-general}
  j_{s}=-\frac{4e}{\hbar} \im \left\{ K \psi^* \nabla_x\psi + \tilde{K}\left[ (\nabla_x\psi)^*\nabla_x^2\psi-\psi^*\nabla_x^3\psi \right] \right\}.
\end{equation}
It is easy to see that the state (\ref{FF-state}) carries zero current. However, by connecting the wire to an external source one can pass a nonzero supercurrent 
through the wire, which means that, in general, thermodynamics of the system should be analyzed taking into account the fixed current constraint, see below.   

It is convenient to introduce the dimensionless order parameter $\tilde{\psi}$ and coordinate $\tilde{x}$ according to
\begin{equation}
\label{dim-less-notations-1}
  \psi=\psi_0\tilde{\psi},\quad x=\xi\tilde{x},
\end{equation}
where $\psi_0=\sqrt{|\alpha|/\beta}$ and the characteristic length $\xi=\sqrt{|K|/|\alpha|}$ is defined similarly to the usual GL correlation length. 
Then we obtain for the free energy density and the supercurrent: 
\begin{equation}
\label{dim-less-notations-2}
  F=\frac{\alpha^2}{\beta}\tilde F,\quad j_s=\frac{4e}{\hbar}\frac{|K|\psi_0^2}{\xi}\,{\tilde j}_s,
\end{equation}
where $\tilde F$ and ${\tilde j}_s$ are dimensionless. Omitting the tildes, we arrive at the dimensionless form of Eq. (\ref{modified-GL-energy-density}):
\begin{equation}
\label{F-dim-less}
	F = -|\psi|^2 + \frac{1}{2}|\psi|^4 - |\nabla_x\psi|^2 + \zeta|\nabla_x^2\psi|^2,
\end{equation}
where 
\begin{equation}
	\zeta=\frac{|\alpha|\tilde{K}}{K^2}>0.
    \label{zeta-def}
\end{equation}
The free energy minimization with respect to $\psi^*$ produces the following modified GL equation:
\begin{equation}
\label{dim-less-GL-eq}
  \zeta \nabla_x^4\psi + \nabla_x^2\psi + \psi(|\psi|^2-1) = 0,   
\end{equation}
while the supercurrent (\ref{supercurrent-general}) in the dimensionless notation takes the form
\begin{equation}
\label{dim-less-j_s}
  j_s = \im \left\{\psi^*\nabla_x\psi - \zeta\left[(\nabla_x\psi)^*\nabla_x^2\psi - \psi^*\nabla_x^3\psi\right] \right\}.
\end{equation}
Note that this last expression can also be obtained by replacing $\nabla_x\to\nabla_x+iA$ in Eq. (\ref{F-dim-less}) and varying the resulting free energy with respect to the dimensionless vector potential $A$ as follows:
\begin{equation}
\label{j_s-var-derivative}
  j_s(x)=-\frac{1}{2}\left.\frac{\delta{\cal F}}{\delta A(x)}\right|_{A=0}.
\end{equation}
The corresponding equations in the usual (non-FFLO) case are recovered if $K>0$ and $\tilde K=0$, \textit{i.e.}, by setting $\zeta = 0$ while simultaneously reversing the signs in front of the second derivative term in 
Eq. (\ref{dim-less-GL-eq}) as well as that of $j_s$. 

An important insight can be obtained by representing the order parameter in the amplitude-phase form, $\psi(x) = \Delta(x) e^{i\theta(x)}$. Then the free energy ${\cal F}$ becomes a functional of $\Delta$, $\theta$, and 
the vector potential $A$, the last two fields entering only via their gauge-invariant combination 
\begin{equation}
\label{v_s-def}
  v_s=\nabla_x\theta+A,
\end{equation}
which is nothing but the dimensionless superfluid velocity. Since, according to Eq. (\ref{j_s-var-derivative}),
\begin{equation}
\label{delta F-delta theta}
  \frac{\delta{\cal F}}{\delta\theta}=-\frac{d}{dx}\left[\frac{\partial F}{\partial(\nabla_x\theta)}-\frac{d}{dx}\frac{\partial F}{\partial(\nabla_x^2\theta)}+...\right]
  =-\frac{d}{dx}\left[\frac{\partial F}{\partial A}-\frac{d}{dx}\frac{\partial F}{\partial(\nabla_x A)}+...\right]=-\frac{d}{dx}\frac{\delta{\cal F}}{\delta A}=2\frac{dj_s}{dx},
\end{equation}  
the minimization of ${\cal F}$ with respect to $\theta$ reproduces the current conservation condition $dj_s/dx=0$. Therefore, in equilibrium we have 
\begin{equation}
\label{constant-current}
  j_s=\mathrm{const}=I.
\end{equation}
This conclusion is completely general and is valid for any form of the GL functional.

We assume that the value of the current $I$ is fixed by an external source. If $I=0$, then $\delta{\cal F}/\delta A=0$, i.e., the zero-current state corresponds to a minimum of the Helmholtz 
free energy with respect to the vector potential. However, if $I\neq 0$, then the 
equilibrium state does not correspond to a minimum of ${\cal F}$. It has long been understood in the usual (non-FFLO) case, that the current-biased superconducting states are obtained by minimizing a different 
thermodynamic potential called the Gibbs free energy.\cite{Tinkham-book,McCumb68} In our case, the dimensionless gauge-invariant Gibbs energy density is given by $G=F+2Iv_s$, therefore 
\begin{equation}
\label{Gibbs-energy-functional}
  {\cal G}={\cal F}+2I\int_0^L (\nabla_x\theta+A)\,dx.
\end{equation}
The current $I$ is regarded as an independent variable, while the phase difference between the ends of the wire becomes a dependent variable and is determined by $I$. The equations for the order parameter obtained 
from ${\cal G}$ and ${\cal F}$ are the same, $\delta{\cal G}/\delta\psi^*=\delta{\cal F}/\delta\psi^*=0$, and it follows from Eqs. (\ref{j_s-var-derivative}) and (\ref{constant-current}) that
$\delta{\cal G}/\delta A=\delta{\cal F}/\delta A+2I=0$. Therefore, the equilibrium current-carrying states indeed correspond to the minima of ${\cal G}$.

\begin{figure}
\includegraphics[width=9cm]{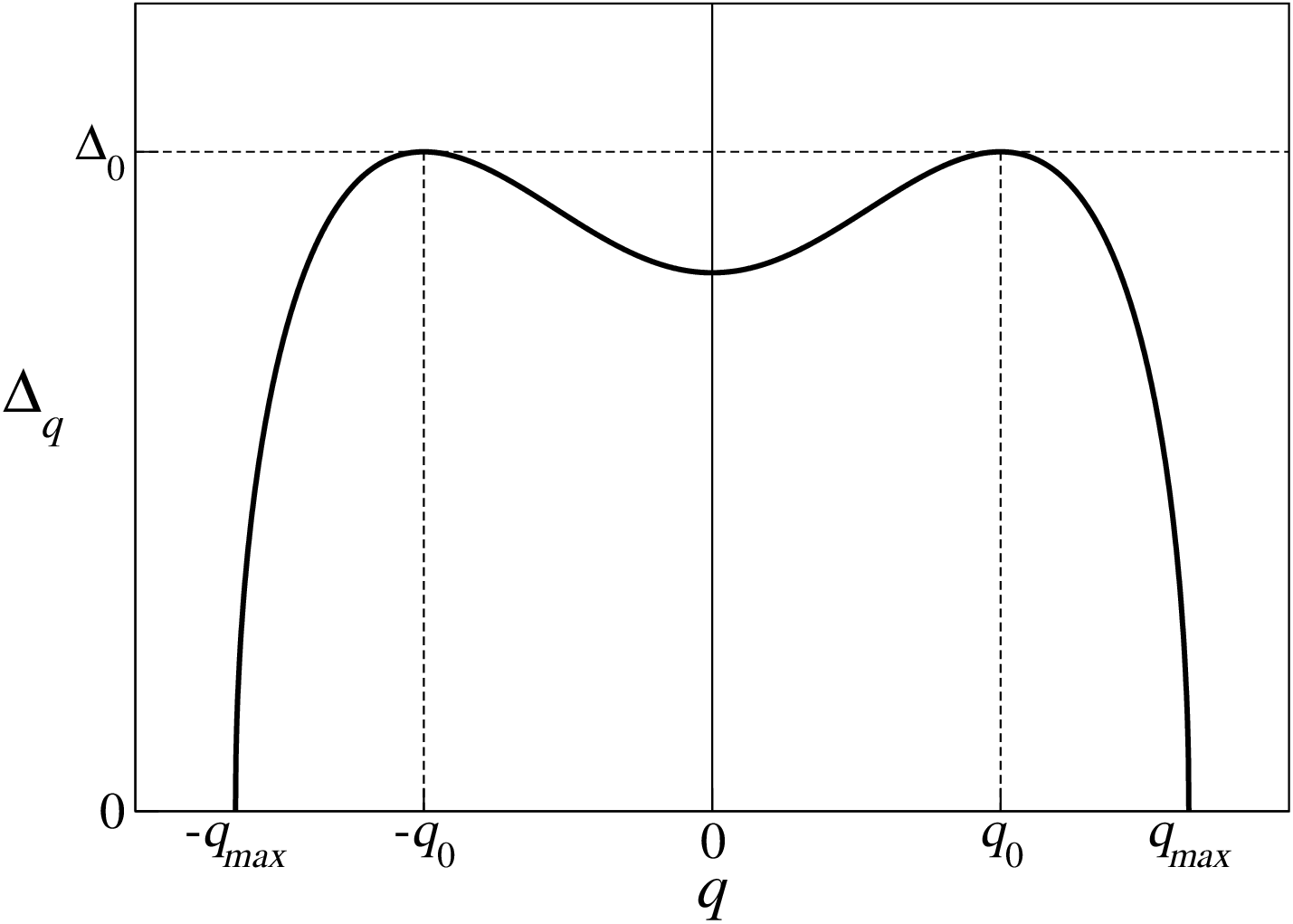}
\caption{The gap magnitude in the FF state as a function of the wavevector $q$. The maximum gap ($\Delta_0=\Delta_{\pm q_0}$) corresponds to the zero-current FF state.}
\label{fig: Delta-q}
\end{figure}

\begin{figure}
\includegraphics[width=9cm]{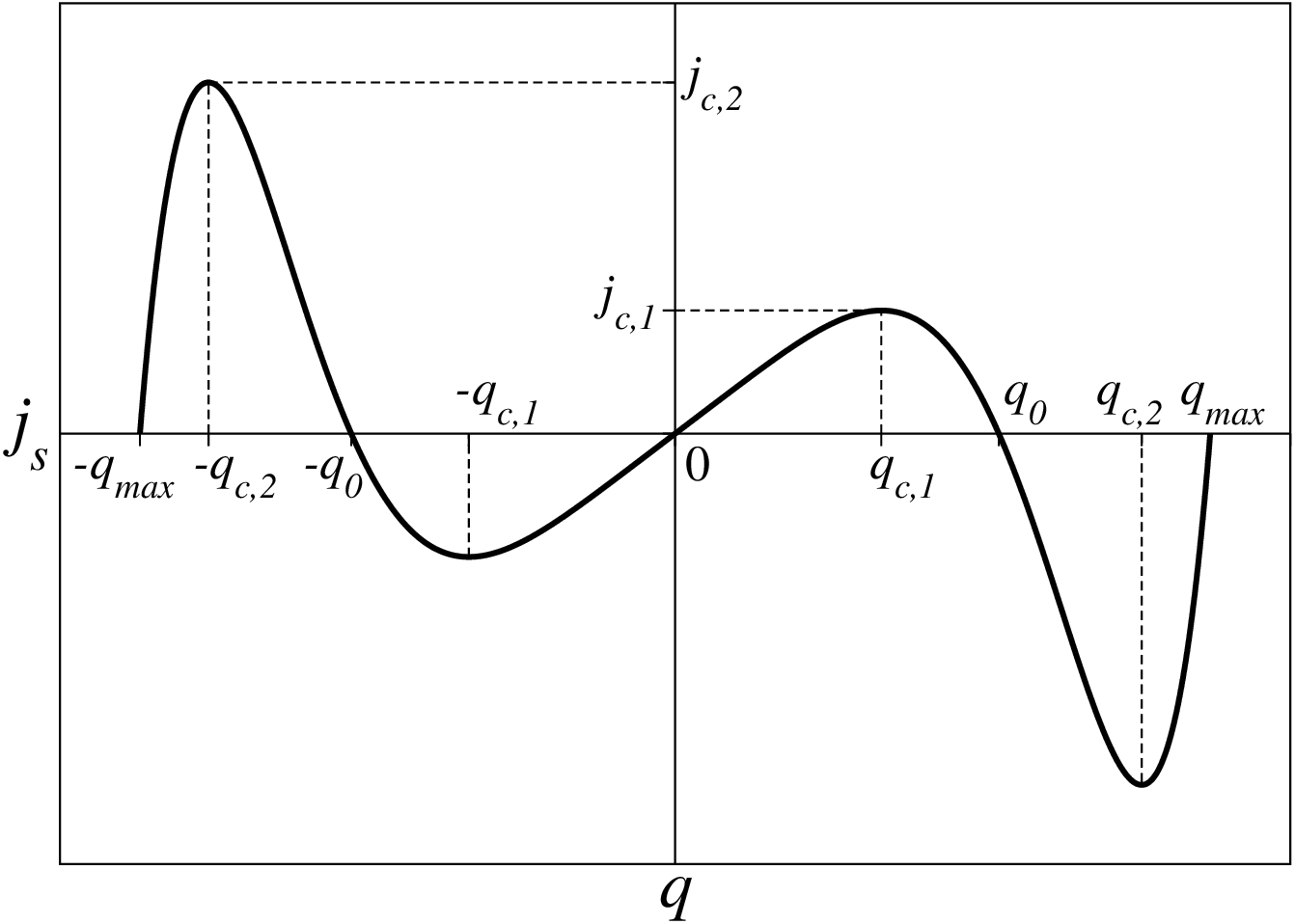}
\caption{The critical values of supercurrent $j_s$ carried by the FF state, see Eq. (\ref{FF-current-q}).}
\label{fig: j_s-q}
\end{figure}

In the gauge $A=0$, which is assumed henceforth, the GL equation takes the form (\ref{dim-less-GL-eq}). It has a simple exact solution 
\begin{equation}
\label{FF-general}
  \psi(x)=\Delta_q e^{iqx},
\end{equation}
with $\Delta_q^2=1+q^2-\zeta q^4$, which describes a current-carrying FF state. The corresponding Gibbs energy density is given by
\begin{equation}
\label{G-FF}
  G(q)=-\frac{1}{2}\Delta_q^4+2Iq,
\end{equation}
while the substitution of the FF solution into Eq. (\ref{dim-less-j_s}) produces the following expression for the supercurrent:
\begin{equation}
\label{FF-current-q}
  j_s(q)=\Delta_q^2(1-2\zeta q^2)q.
\end{equation}
The solution (\ref{FF-general}) exists only at 
$$
  |q|\leq q_{max} = \sqrt{\frac{1+\sqrt{1+4\zeta}}{2\zeta}}.
$$
Plots of $\Delta_q$ and $j_s(q)$ are shown in Figs. \ref{fig: Delta-q} and \ref{fig: j_s-q}. It should be noted that our results do not qualitatively depend on the value of $\zeta$, and we use $\zeta=0.5$ in all plots. 

We see that the way the FF superconductor responds to an externally applied current is very different from the textbook case of a BCS superconductor, see, \textit{e.g.}, Ref. \onlinecite{Tinkham-book}. 
The supercurrent has two critical values, given by
\begin{equation}
\label{j_c-defs}
  j_{c,1}=\pm j_s(\pm q_{c,1}),\quad j_{c,2}=\mp j_s(\pm q_{c,2}),
\end{equation}
where
\begin{equation}
\label{q_c-12}
  q_{c,1}^2 = \frac{\sqrt{11+28\zeta}}{7\zeta}\cos\left(\gamma-\frac{2\pi}{3}\right)+\frac{5}{14\zeta},\quad 
  q_{c,2}^2 = \frac{\sqrt{11+28\zeta}}{7\zeta}\cos\gamma+\frac{5}{14\zeta},
\end{equation}
and
$$
	\gamma = \frac{1}{3}\arccos\frac{4(5+28\zeta)}{(11+28\zeta)^{3/2}}.
$$
The upper critical current $j_{c,2}$ corresponds to the maximum current that can be passed through the system, an analog of the BCS depairing current. One can show that $j_{c,1} < j_{c,2}$ for all $\zeta$. 
In the limit $\zeta\gg 1$ [according to Eq. (\ref{zeta-def}), this is realized at $K\to-0$, \textit{i.e.}, in the vicinity of the FF tricritical point], Eq. (\ref{q_c-12}) yields $q_{c,1}\simeq(6\zeta)^{-1/2}$ and 
$q_{c,2}\simeq(7\zeta/3)^{-1/4}$.

According to Eq. (\ref{constant-current}), the wavevector $q$ of the FF state depends on the applied current and is found by solving the equation $j_s(q)=I$. 
At zero current, there are two degenerate states with $q=\pm q_0$, where 
\begin{equation}
	q_0 = \frac{1}{\sqrt{2\zeta}},
    \label{q_0}
\end{equation}
cf. Eq. (\ref{FFLO-wavevector}), with the gap magnitude $\Delta_0\equiv\Delta_{\pm q_0}= \sqrt{1 + 1/4\zeta}$. In general, as evident from Fig. \ref{fig: j_s-q}, the equation $j_s(q)=I$ has two solutions at $j_{c,1}<|I|<j_{c,2}$ 
and four solutions at $|I|<j_{c,1}$, which suggests that there are multiple superconducting states able to sustain the given current. However, not all these states are stable, see the next subsection.

\subsection{Stability analysis}
\label{sec: SC-stability}

The local stability of the current-carrying FF states found above can be investigated using the approach similar to Ref. \onlinecite{LA67}. 
We consider the solution (\ref{FF-general}) subject to a small complex-valued perturbation $\delta\Delta$ as follows: 
\begin{equation}
\label{FF-perturbed}
	\psi(x) = \left[\Delta_q + \delta\Delta(x)\right]e^{iqx}.
\end{equation}
Substitution of this into Eqs. (\ref{Gibbs-energy-functional}) and (\ref{F-dim-less}) yields $G=G_0+\delta G$, where $G_0$ is the Gibbs energy density of the unperturbed FF state, see Eq. (\ref{G-FF}), and
$\delta G$ is the correction, in which we retain only the terms quadratic in $\delta\Delta$ and its derivatives.

\begin{figure}
\includegraphics[width=9cm]{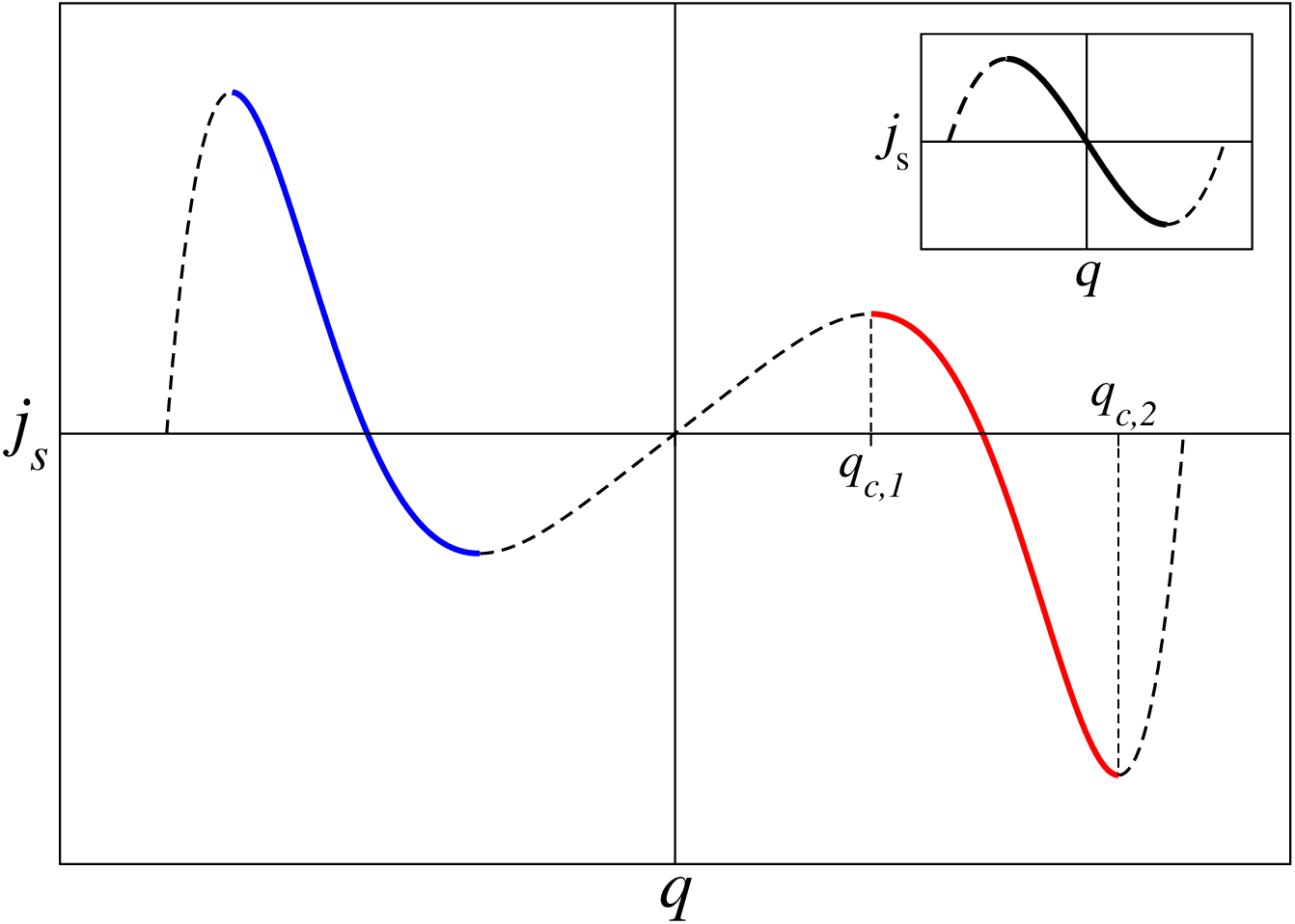}
\caption{(Color online) Stable and unstable branches of the current-carrying FF states. The solid regions of the curve correspond to the locally stable solutions, 
	    while the dashed regions correspond to unstable solutions. The usual BCS case is shown in the inset for comparison (recall that the electron charge is $-e$, therefore $j_s$ is negative at $q>0$).}
\label{fig: j_s-stability}
\end{figure}

Expressing the order parameter deformation in terms of its real and imaginary parts, $\delta\Delta(x)=f_1(x)+if_2(x)$, we obtain: $\delta G = \bm{f}^\top \hat{\cal L}_q \bm{f}$,
where $\bm{f}=(f_1,f_2)^\top$ and
$$
	\hat{\cal L}_q = 
    \begin{pmatrix}
    	\zeta\nabla_{x}^4 + (1-6\zeta q^2)\nabla_{x}^2 + 2\Delta_q^2 & - 4\zeta q\nabla_{x}^3 + 2q(2\zeta q^2 - 1)\nabla_{x} \\ \\
        4\zeta q\nabla_{x}^3 - 2q(2\zeta q^2 - 1)\nabla_{x} & \zeta\nabla_{x}^4 + (1-6\zeta q^2)\nabla_{x}^2
    \end{pmatrix}.
$$
The eigenfunctions of the matrix operator $\hat{\cal L}_q$ are plane waves $\propto e^{ikx}$, with the eigenvalues given by  
\begin{equation}
	\lambda^{\pm}_q(k) = \zeta k^4 + (6\zeta q^2 -1)k^2 + \Delta_q^2 \pm \sqrt{\Delta_q^4 + 4q^2k^2[2\zeta(q^2 + k^2) -1]^2},
    \label{L-eigenvalues}
\end{equation}
at given $q$. Note that $\lambda^-_q(0)=0$, which describes the Goldstone mode corresponding to a uniform phase rotation of the FF state. 
The Gibbs energy has a stable minimum if $\lambda^\pm_q(k)>0$ for all $k\neq 0$. A straightforward inspection of Eq. (\ref{L-eigenvalues}) shows that this last condition is satisfied if 
\begin{equation}
\label{local-stability-condition}
	q_{c,1}\leq |q| \leq q_{c,2},
\end{equation}
i.e., the regions of local stability of the current-carrying states are bounded by the critical points of the supercurrent $j_s(q)$, see Eq. (\ref{q_c-12}). 
This is shown in Fig. \ref{fig: j_s-stability}, along with the corresponding plot in the usual BCS case.

The stability condition (\ref{local-stability-condition}) can also be understood using a simple general argument, which works for an arbitrary structure of the gradient energy. 
Assuming that the equilibrium current-carrying solution to the GL equations as well as its small perturbations are represented by plane waves, we consider the order parameter of the form $\psi(x) = \Delta e^{iqx}$, where 
$\Delta$ and $q$ are real constants. Then the corresponding Gibbs energy density becomes a function of $\Delta$ and $q$, so that 
\begin{equation}
	G(\Delta, q) = F(\Delta, q) + 2Iq,
    \label{G-Delta-q}
\end{equation}
see Eq. (\ref{Gibbs-energy-functional}). We do not specify here the dependence of the free energy on $\Delta$ and $q$. 
Minimization of Eq. (\ref{G-Delta-q}) at given $I$ produces two equations:
\begin{equation}
\label{dG-dDelta}
  \frac{\partial G}{\partial \Delta}=0,
\end{equation}
which yields the magnitude $\Delta_q$ of the solution, and $\partial G/\partial q=0$, which is used to find the wavevector $q(I)$.

The stability of the equilibrium state is determined by the signature of the matrix of the second derivatives of $G$ evaluated at $\Delta=\Delta_q$ and $q=q(I)$ (the Hessian matrix). Namely, the solution becomes unstable 
when the determinant of this matrix changes sign. The Hessian determinant is given by
\begin{equation}
  D(q)=G_{\Delta\Delta}G_{qq}-G_{\Delta q}^2,
    \label{Hessian-det-general}
\end{equation}
where $G_{XY}(q)=(\partial^2G/\partial X\partial Y)|_{\Delta=\Delta_q}$, which can be calculated as follows. From the expression for the supercurrent,
$$
  j_s(q)=-\frac{1}{2}\left.\frac{\partial F}{\partial q}\right|_{\Delta=\Delta_q},
$$
and Eq. (\ref{G-Delta-q}) we obtain:
\begin{equation}
\label{dj_s-dq}
  \frac{dj_s}{dq} = -\frac{1}{2}\left(G_{\Delta q}\frac{d\Delta_q}{dq}+G_{qq}\right).
\end{equation}
On the other hand, it follows from Eq. (\ref{dG-dDelta}) that
\begin{equation}
\label{dDelta-equation}
  G_{\Delta\Delta}\frac{d\Delta_q}{dq}+G_{\Delta q}=0.
\end{equation}
Using Eqs. (\ref{dj_s-dq}) and (\ref{dDelta-equation}), the Hessian determinant (\ref{Hessian-det-general}) can be written in the following form:
\begin{equation}
\label{Hessian-det-final}
  D(q)=-2G_{\Delta\Delta}\frac{dj_s}{dq}=-2\left(\frac{\partial^2F}{\partial\Delta^2}\right)_{\Delta=\Delta_q}\frac{dj_s}{dq}.
\end{equation}
At $|q|<q_{max}$, this last expression passes through zero only when $dj_s(q)/dq=0$, at which point a plane-wave like superconducting state becomes unstable. Thus the condition (\ref{local-stability-condition}) is reproduced.

\subsection{Electric current in the FF state}
\label{sec: metastable}

\begin{figure}
\includegraphics[width=9cm]{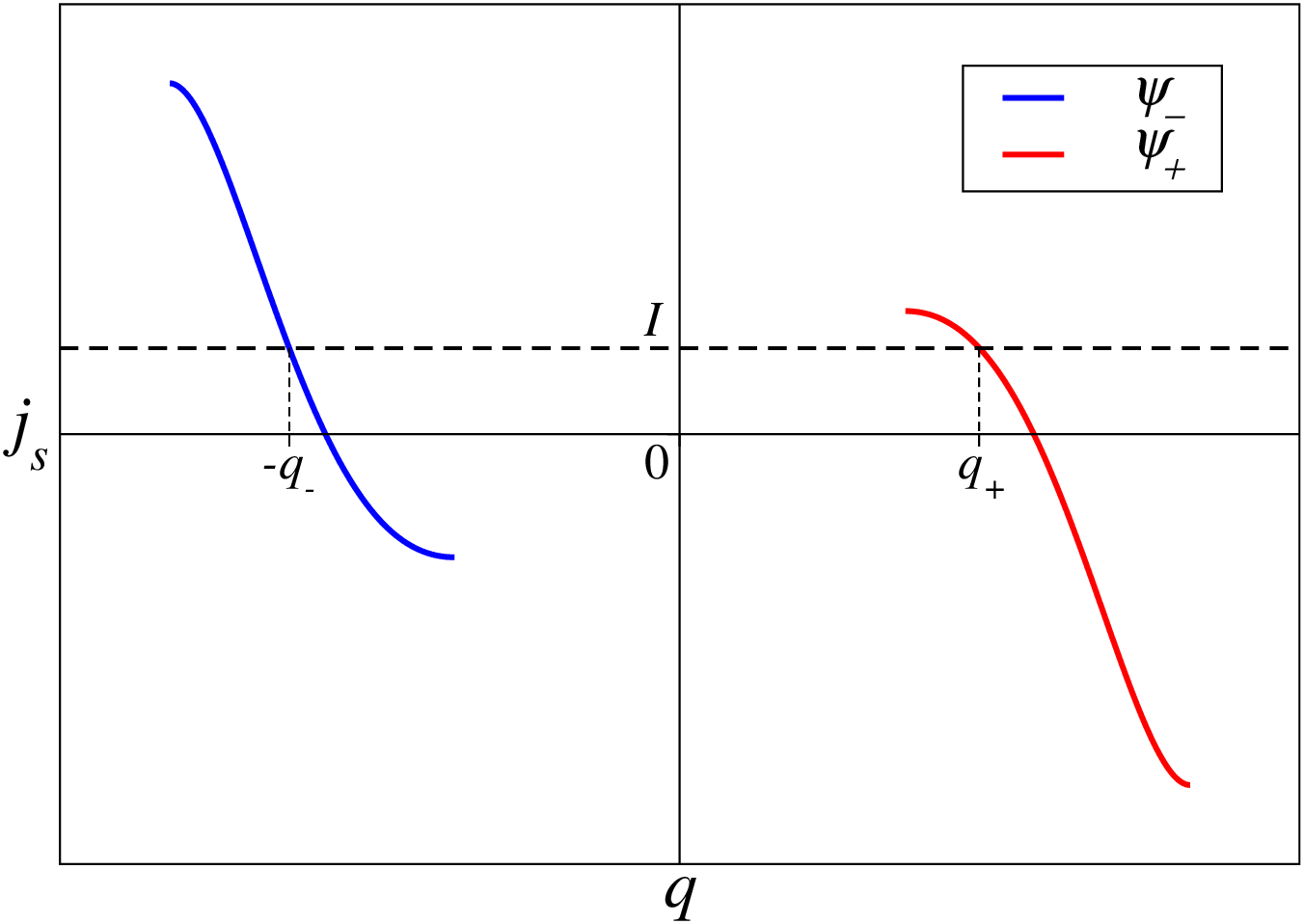}
\caption{(Color online) The locally stable FF states, Eqs. (\ref{psi-plus}) and (\ref{psi-minus}). }
\label{fig: qp-qm}
\end{figure}

\begin{figure}
\includegraphics[width=9cm]{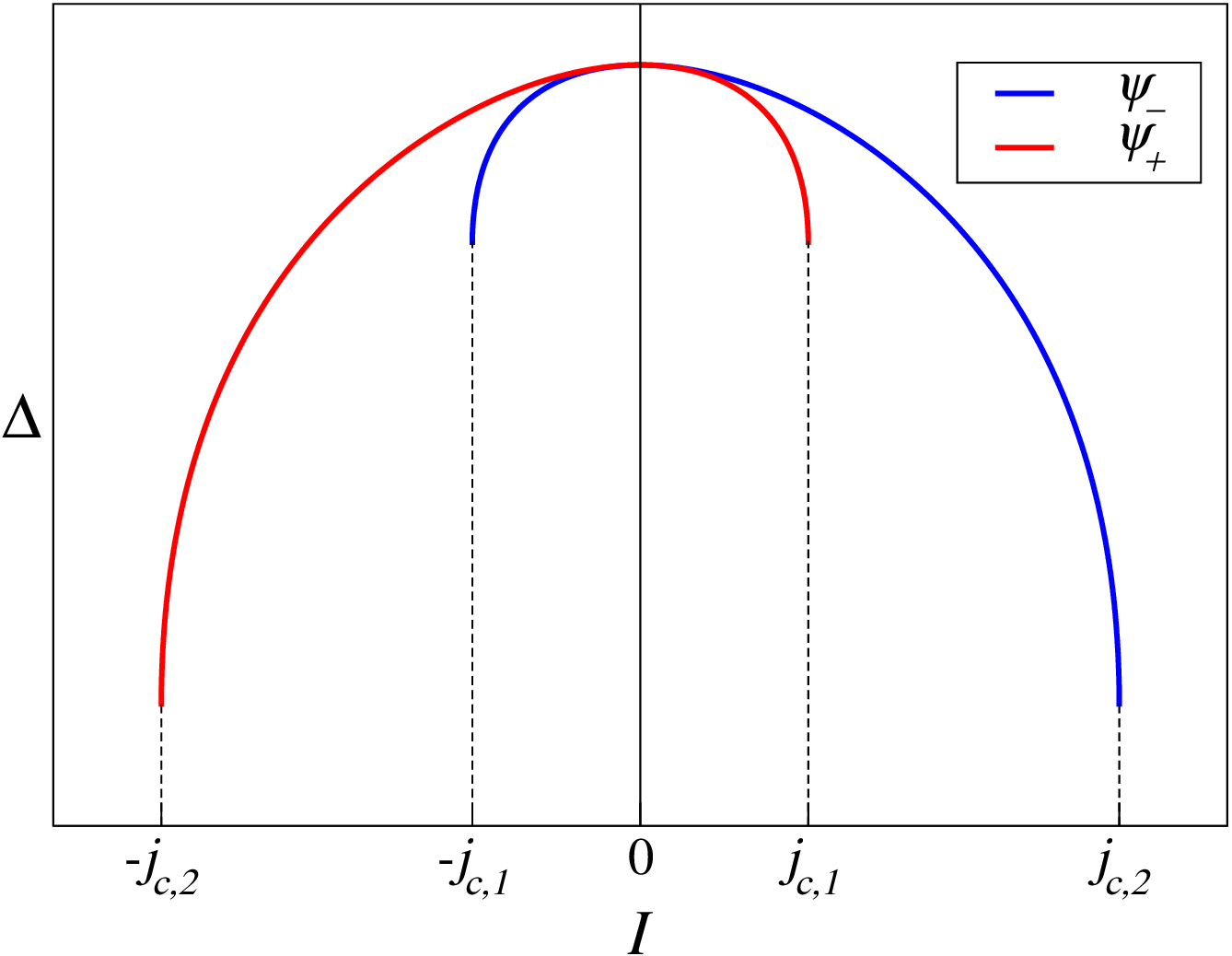}
\caption{(Color online) The gap magnitudes of the FF states (\ref{psi-plus}) and (\ref{psi-minus}), as functions of the applied current.}
\label{fig: gap-vs-I}
\end{figure}

At small applied currents, $|I|<j_{c,1}$, there exist two locally stable FF states:
\begin{equation}
\label{psi-plus}
  \psi_+(x)=\Delta_+e^{iq_+x}
\end{equation}
and
\begin{equation}
\label{psi-minus}
  \psi_-(x)=\Delta_-e^{-iq_-x},
\end{equation}
where $\Delta_+\equiv\Delta_{q_+}$, $\Delta_-\equiv\Delta_{-q_-}=\Delta_{q_-}$, and $q_\pm$ are found by solving the equation $j_s(q)=I$, see Fig. \ref{fig: qp-qm}. At $j_{c,1}<I<j_{c,2}$, 
the current is carried by the state (\ref{psi-minus}), while at $-j_{c,2}<I<-j_{c,1}$, the current is carried by the state (\ref{psi-plus}). 
The dependence of the gap magnitude of the states (\ref{psi-plus}) and (\ref{psi-minus}) on $I$ is shown in Fig. \ref{fig: gap-vs-I}. 

\begin{figure}
\includegraphics[width=9cm]{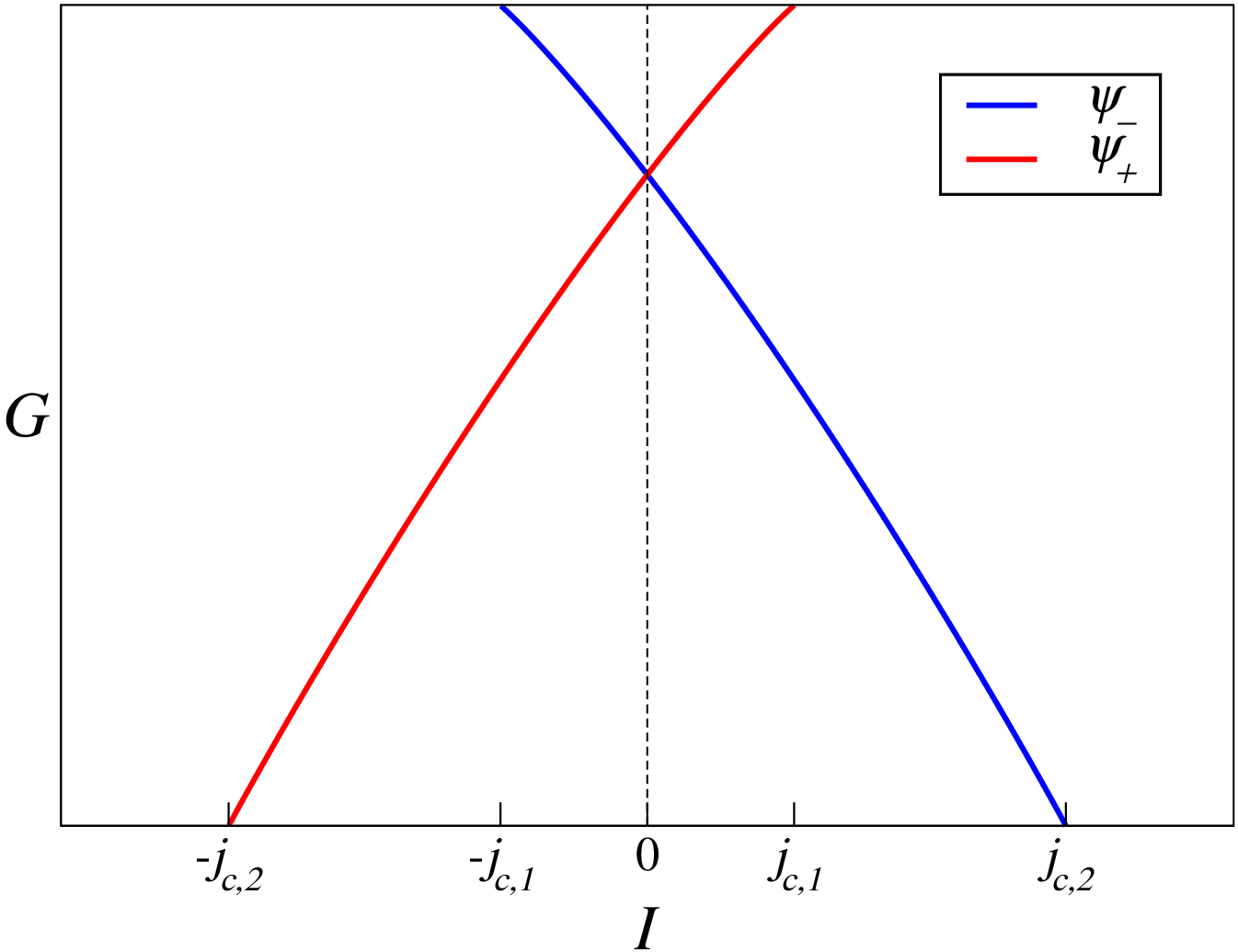}
\caption{(Color online) The Gibbs free energies of the FF states (\ref{psi-plus}) and (\ref{psi-minus}), as functions of the applied current.}
\label{fig: Gibbs energies}
\end{figure}

Calculating the Gibbs free energy, Eq. (\ref{G-FF}), for the states (\ref{psi-plus}) and (\ref{psi-minus}) and introducing the notation 
$$
  \Delta G=G(q_+)-G(-q_-),
$$ 
we arrive at the following picture. At $I=0$, the two FF states have the same Gibbs energy (which is equal to the free energy). At $0<I\leq j_{c,1}$, we have $\Delta G>0$, therefore $\psi_+$ is metastable. 
At $-j_{c,1}\leq I<0$, we have $\Delta G<0$, therefore $\psi_-$ is metastable. This is shown in Fig. \ref{fig: Gibbs energies}. We see from Fig. \ref{fig: gap-vs-I} that the state 
with the larger gap magnitude is always more stable. At small currents, we have $q_\pm\simeq q_0\mp I/2\Delta_0^2$ and the energy difference is given by $\Delta G\simeq 4q_0I$.

The existence of two distinct superconducting states available to conduct a given current can lead to a characteristic ``branch switching'', which might be observed in experiment. 
Suppose that at zero applied current the superconducting wire is in the state $\psi_+$ (the possibility of a coexistence of $\psi_+$ and $\psi_-$ separated by a domain wall will be discussed in the next subsection). 
Increasing the current, \textit{i.e.}, at $0<I\leq j_{c,1}$, this state becomes metastable and can decay into $\psi_-$ by a first-order transition.
During the transition, a localized nucleus of $\psi_-$ appears as a fluctuation, which would eventually grow to fill the whole system. The probability of this activation process is determined by the energy barrier separating
$\psi_+$ and $\psi_-$. While a quantitative theory, in particular, calculating the height of this barrier, is beyond the scope of the present work, 
we note that the mechanism of the switching between the two current-carrying states may be qualitatively similar 
to the formation of a phase slip center in a non-FFLO superconducting wire.\cite{LA67,McCumb68,MCH70}

\subsection{Domain walls}
\label{sec: DW}

In addition to the single-plane wave states (\ref{psi-plus}) and (\ref{psi-minus}), the GL equations have more complicated solutions with a finite energy. For instance, the FF states at zero current, 
$\psi_\pm(x)=\Delta_0e^{\pm iq_0x}$ are degenerate and can therefore be separated by a stationary domain wall. To find the corresponding solution, we substitute the general form of the 
order parameter $\psi(x) = \Delta(x) e^{i\theta(x)}$ in Eqs. (\ref{dim-less-GL-eq}) and (\ref{dim-less-j_s}), and obtain:
\begin{equation}
\label{GL-eq-Delta-theta}
  \zeta (\nabla_x+iv_s)^4\Delta+(\nabla_x+iv_s)^2\Delta+(\Delta^2-1)\Delta=0,
\end{equation}
where $v_s=\nabla_x\theta$, see Eq. (\ref{v_s-def}), and
\begin{equation}
\label{j_s-Delta-theta}
  j_s=\Delta^2v_s+\zeta\left[\Delta^2\nabla_x^2v_s-2\Delta^2v_s^3+4\Delta(\nabla_x^2\Delta)v_s+2\Delta(\nabla_x\Delta)(\nabla_xv_s)-2(\nabla_x\Delta)^2v_s\right].
\end{equation}
The real and imaginary parts of Eq. (\ref{GL-eq-Delta-theta}) are given by
\begin{equation}
\label{re-part-GL-eq}
  \zeta(\hat R_1^2-\hat R_2^2)\Delta+\hat R_1\Delta+(\Delta^2-1)\Delta=0
\end{equation}
and
\begin{equation}
\label{im-part-GL-eq}
  \zeta\{\hat R_1,\hat R_2\}\Delta+\hat R_2\Delta=0,
\end{equation}
respectively. Here $\hat R_1=\nabla_x^2-v_s^2$, $\hat R_2=\{\nabla_x,v_s\}$, and the curly brackets denote the anticommutator of two operators. It is straightforward to check that $\nabla_xj_s$ is equal to the left-hand side of
Eq. (\ref{im-part-GL-eq}) multiplied by $\Delta$, therefore $\nabla_xj_s=0$ at all $x$ and the current conservation, Eq. (\ref{constant-current}), is reproduced.

\begin{figure}
\includegraphics[width=9cm]{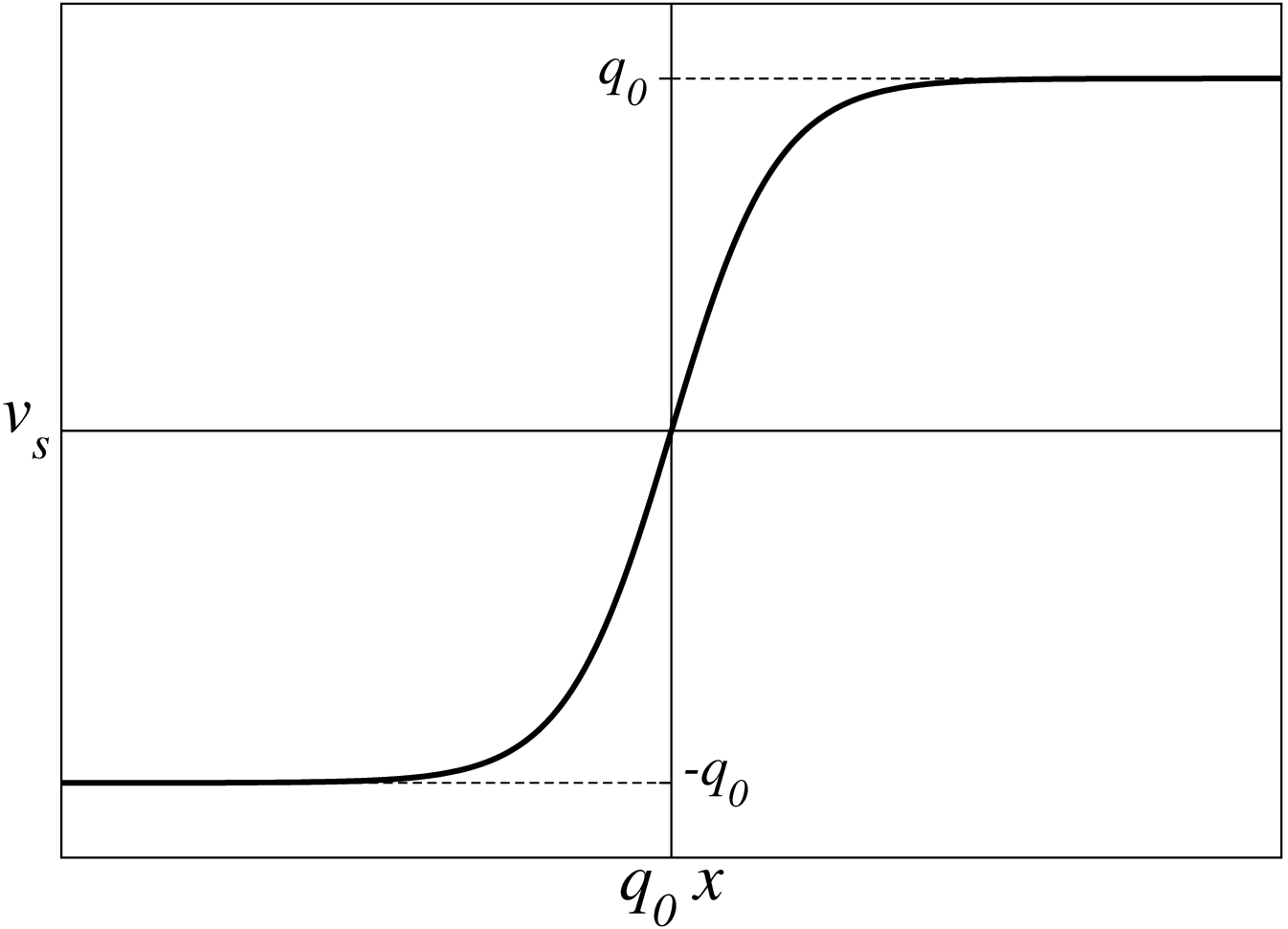}
\caption{Domain wall connecting the degenerate FF states (\ref{psi-plus}) and (\ref{psi-minus}) at zero applied current. }
\label{fig: DW}
\end{figure}

Possible nonuniform textures of the order parameter at given $I$ can be found by solving the coupled equations (\ref{re-part-GL-eq}) and (\ref{j_s-Delta-theta}), with the constraint $j_s=I$. Focusing on the domain wall 
state at zero current, we impose the boundary conditions $\Delta(x \rightarrow \pm \infty) = \Delta_0$ and $v_s(x \rightarrow \pm \infty) = \pm q_0$.
We seek an approximate solution with $\Delta(x) = \Delta_0$ everywhere. Then, Eq. (\ref{j_s-Delta-theta}) yields the following equation for the superfluid velocity:
$$
  \zeta \frac{d^2v_s}{dx^2} - 2\zeta v_s^3 + v_s = 0,
$$
which has the first integral of the form
\begin{equation}
	\frac{1}{2}\left( \frac{dv_s}{dx}\right)^2 - \frac{1}{2}v_s^4 + \frac{1}{2 \zeta}v_s^2 = C.
	\label{v_s-1st-integral}
\end{equation}
Here $C = 1/8 \zeta^2$, according to the boundary conditions. From Eq. (\ref{v_s-1st-integral}) we obtain:
\begin{equation}
  v_s(x) = \frac{1}{\sqrt{2 \zeta}} \tanh \left( \frac{x}{\sqrt{2 \zeta}} \right),
    \label{DW-v_s-solution}
\end{equation}
therefore $\theta(x) = \ln[2\cosh(x/\sqrt{2\zeta})]$. The domain wall profile is shown in Fig. \ref{fig: DW}.

We now calculate the energy cost of the domain wall. Observing that at zero current $G=F$ and using Eq. (\ref{v_s-1st-integral}), we obtain:
$$
	F = -\frac{1}{2}\Delta_0^4+2\zeta\Delta_0^2\left(\frac{dv_s}{dx}\right)^2.
$$
The last term here represents the extra gradient energy due to the order parameter variation. Substituting Eq. (\ref{DW-v_s-solution}), we arrive at the following expression for the domain wall energy:
\begin{equation}
\label{DW-energy}
  \epsilon_{DW}=2\zeta\Delta_0^2\int_{-\infty}^\infty dx\left(\frac{dv_s}{dx}\right)^2=\frac{4}{3 \sqrt{2 \zeta}} \left(1 + \frac{1}{4 \zeta} \right).
\end{equation}
It should be noted that the constant-magnitude approximation is valid only at $\zeta\gg 1$, which corresponds to a ``wide'' domain wall. 
One can see from Eq. (\ref{re-part-GL-eq}) that in this limit $\Delta(x)=\Delta_0+\delta\Delta(x)$, where $\delta\Delta\propto{\cal O}(\zeta^{-1})$.

At $I\neq 0$, the FF states (\ref{psi-plus}) and (\ref{psi-minus}) are no longer degenerate. A domain wall between these states is still possible, but it cannot exist as a stationary defect and will move towards the state 
with the lower Gibbs energy. According to Fig. \ref{fig: Gibbs energies}, it will move towards $\psi_-$ if $I>0$, and towards $\psi_+$ if $I<0$. The domain wall motion may result in a nonzero voltage across the wire.

\subsection{FF state in a ring}
\label{sec: FF-ring}

We have shown above that, despite the spatial modulation of the order parameter phase, an FF superconducting wire carries zero supercurrent, unless it is connected to an external current source. 
In this subsection, we consider a wire subject to periodic boundary conditions, \textit{i.e.}, closed into a loop, and argue that the FF state in this geometry can generate a spontaneous supercurrent. 
This system also exhibits some unusual properties due to superconducting fluctuations, see Ref. \onlinecite{ZZ09}. 

The properties of the FF state in a ring of radius $R$ are obtained by minimizing the modified Helmholtz free energy of the form 
\begin{equation}
\label{F-dim-less-ring}
	F = -|\psi|^2 + \frac{1}{2}|\psi|^4 - \frac{1}{M^2}\left|\nabla_\phi\psi\right|^2 + \frac{\zeta}{M^4}\left|\nabla_\phi^2\psi\right|^2,
\end{equation}
cf. Eq. (\ref{F-dim-less}). Here $0\leq\phi<2\pi$ is an angular coordinate along the ring, $M=R/\xi$, and we used the same dimensionless notations as above, see Eqs. (\ref{dim-less-notations-1}) and (\ref{dim-less-notations-2}). 
The nonlinear GL equation that follows from Eq. (\ref{F-dim-less-ring}) has an exact solution $\psi_m(\phi)=\Delta_m e^{im\phi}$, where $m=0,\pm 1,\pm 2,...$, and 
\begin{equation}
\label{Delta-m}
  \Delta_m^2=1+\frac{m^2}{M^2}-\zeta\frac{m^4}{M^4}.
\end{equation}
This superconducting state has the free energy density $F_m=-\Delta_m^4/2$ and carries the supercurrent
\begin{equation}
\label{j_s-ring-m}
  j_{s,m}=\Delta_m^2\left(1-2\zeta\frac{m^2}{M^2}\right)\frac{m}{M},
\end{equation}
cf. Eq. (\ref{FF-current-q}). The results obtained earlier in this section for an infinite wire are recovered by taking the limit $M\to\infty$, in which $q=m/M$ becomes a continuous variable. 
Note that we have neglected the magnetic field energy created by the supercurrent due to the self-inductance of the ring. In the presence of a perpendicular (Aharonov-Bohm) magnetic flux $\Phi$,
either self-induced or external, Eqs. (\ref{Delta-m}) and (\ref{j_s-ring-m}) would be modified by replacing $m\to m+\nu$, where $\nu=\Phi/\Phi_0$ and $\Phi_0=\pi\hbar c/e$ is the magnetic flux quantum.

\begin{figure}
\includegraphics[width=9cm]{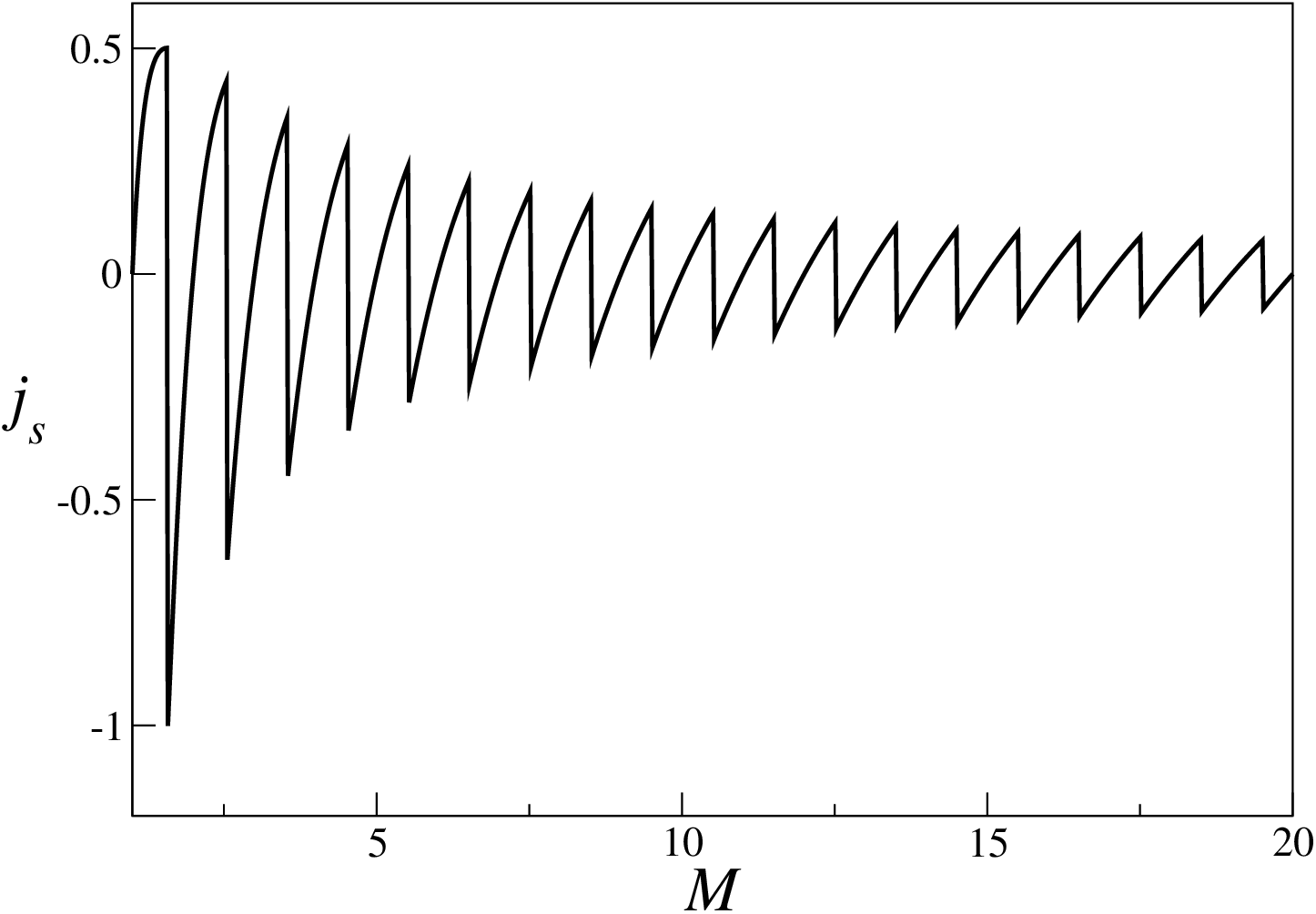}
\caption{The spontaneous supercurrent generated by the FF state in a ring, as a function of the radius of the ring ($M=R/\xi$).}
\label{fig: j_s-vs-M}
\end{figure}

At given $M$, the free energy has to be minimized to determine the optimal phase winding number, which is then substituted in Eq. (\ref{j_s-ring-m}). 
It is easy to see that, remarkably, the supercurrent is nonzero, in general. This is due to the fact that, in contrast to the non-FFLO case, the minimum of the free energy (or the maximum of the gap) 
does not occur at $m=0$. Instead, $F_m$ has a minimum at nonzero values of the winding number, $m=\pm m_0$, which do not necessarily result
in the vanishing of the expression (\ref{j_s-ring-m}). The minimum is degenerate, with the states $\psi_{m_0}$ and $\psi_{-m_0}$ carrying opposite supercurrents of the same magnitude.
In Fig. \ref{fig: j_s-vs-M}, we plotted the spontaneous supercurrent for the state $\psi_{m_0}$. The magnitude of the current decreases as the size of the ring increases and the continuous
limit is approached.

\section{Nonequilibrium properties}
\label{sec: TDGL}

In this section we develop the time-dependent Ginzburg-Landau (TDGL) formalism for the FFLO superconductors and discuss some of its applications. 
The TDGL equation can be obtained phenomenologically by assuming a relaxational dynamics of the order parameter driven out of an equilibrium state and has the following form:\cite{Kopnin-book} 
\begin{equation}
	-\Gamma \frac{\partial \psi}{\partial t} = \frac{\delta{\cal F}}{\delta \psi^*},
    \label{TDGL-general}
\end{equation}
where $\Gamma > 0$ is a real constant. Note that, although the equilibrium in the presence of an external current corresponds to a minimum of the Gibbs free energy ${\cal G}$, instead of the Helmholtz free energy ${\cal F}$,
the variational derivatives of the two thermodynamic potentials with respect to $\psi^*$ are the same, which allows us to use the TDGL equation in the form (\ref{TDGL-general}). 
In the standard fashion, the formalism is made explicitly gauge-invariant by including the electric scalar potential $\varphi(x,t)$ as follows:
\begin{equation}
	-\Gamma \left( \frac{\partial \psi}{\partial t} - i \frac{2e}{\hbar}\varphi \psi \right) = \frac{\delta{\cal F}}{\delta \psi^*}.
	\label{TDGL-gauge-inv}
\end{equation}
The total current density $j$ now comprises of both the supercurrent $j_{s}$ as well as the normal contribution $j_n$. The latter is given by Ohm's law:
\begin{equation}
	j_n= -\sigma_n\frac{\partial \varphi}{\partial x},
	\label{Ohms-law}
\end{equation}
where $\sigma_n$ is the normal state conductivity. Using Eqs. (\ref{modified-GL-energy-density}) and (\ref{supercurrent-general}), we arrive at the following system of equations, which determine the time evolution of both 
the order parameter and the electric field in an FFLO superconducting wire: 
\begin{eqnarray}
    && -\Gamma \left( \frac{\partial \psi}{\partial t} - i \frac{2e}{\hbar}\varphi \psi \right) = \alpha \psi + \beta |\psi|^2 \psi - K\frac{\partial^2 \psi}{\partial x^2} + \tilde{K}\frac{\partial^4 \psi}{\partial x^4}, 
    \label{TDGL-eq-modified} \\
    && j  = -\sigma_n \frac{\partial \varphi}{\partial x} -\frac{4e}{\hbar} \im \left[ K\psi^* \frac{\partial \psi}{\partial x} + \tilde{K} \left(\frac{\partial \psi^*}{\partial x} 
    \frac{\partial^2 \psi}{\partial x^2} - \psi^* \frac{\partial^3 \psi}{\partial x^3} \right)  \right],
    \label{total-j-modified}
\end{eqnarray}
with $K<0$ and $\tilde K>0$. The corresponding equations in the non-FFLO case are obtained by setting $K>0$ and $\tilde K=0$. 

In the nonequilibrium theory, there is an additional characteristic length -- the electric field penetration depth. To understand this, we consider the static case in a half-infinite geometry, 
with the superconductor at $x>0$. Multiplying Eq. (\ref{TDGL-gauge-inv}) by $\psi^*$ and taking the imaginary part, we obtain:
\begin{equation}
\label{Im-TDGL-static}
  \frac{2e\Gamma}{\hbar}|\psi|^2\varphi=\im\left(\psi^*\frac{\delta{\cal F}}{\delta \psi^*}\right).
\end{equation}
Using the amplitude-phase representation of the order parameter, $\psi=|\psi|e^{i\theta}$, it is easy to show that
\begin{equation}
\label{useful-identity}
  \im\left(\psi^*\frac{\delta{\cal F}}{\delta \psi^*}\right)=\frac{1}{2}\frac{\delta{\cal F}}{\delta\theta}=\frac{\hbar}{4e}\frac{\partial j_s}{\partial x}.
\end{equation}
The last equality here is the dimensional analog of Eq. (\ref{delta F-delta theta}). The current conservation implies that $\partial j_s/\partial x = -\partial j_n/\partial x$, which, taken with Eq. (\ref{Ohms-law}),
leads to the following equation for the scalar potential in the superconductor:
$$
  \frac{\partial^2 \varphi}{\partial x^2} = \frac{8e^2\Gamma}{\hbar^2\sigma_n}|\psi|^2\varphi.
$$
Assuming that $|\psi|=\psi_0$ is constant, as is the case in the FF state, the solution is given by $\varphi(x) \propto e^{-x/l_E}$, where
\begin{equation}
\label{l_E-def}
	l_{E}=\sqrt{\frac{\hbar^2\sigma_n}{8e^2\Gamma\psi_0^2}}
\end{equation}
may be interpreted as the electric field penetration depth or, equivalently, the length scale over which the normal current into the FF superconductor is converted into supercurrent.\cite{Tinkham-book,Kopnin-book} 
Note that Eq. (\ref{l_E-def}) has exactly the same form as in the usual BCS case, which is not surprising since the identity (\ref{useful-identity}) does not depend on the structure of the GL gradient terms. 

One can represent Eqs. (\ref{TDGL-eq-modified}) and (\ref{total-j-modified}) in a dimensionless form by introducing, in addition to the quantities defined in Eqs. (\ref{dim-less-notations-1}) 
and (\ref{dim-less-notations-2}), also the dimensionless time variable $\tilde t$, the scalar potential $\tilde{\varphi}$, and the total current $I$ as follows:
$$
  t = \frac{\hbar^2}{8e^2}\frac{\sigma_n}{|K|\psi_0^2}\tilde{t},\quad \varphi = \frac{4e}{\hbar}\frac{|K|\psi_0^2}{\sigma_n}\tilde{\varphi},\quad j=\frac{4e}{\hbar}\frac{|K|\psi_0^2}{\xi}\,I.
$$ 
Dropping the tildes, we obtain:
\begin{eqnarray}
    && -u\left( \frac{\partial \psi}{\partial t} - i\varphi \psi \right) = \zeta \frac{\partial^4 \psi}{\partial x^4} + \frac{\partial^2 \psi}{\partial x^2} + \psi(|\psi|^2 -1), 
    \label{dim-less-TDGL-eq} \\
    && I = -\frac{\partial \varphi}{\partial x} +\im \left[\psi^* \frac{\partial \psi}{\partial x} - \zeta \left(\frac{\partial \psi^*}{\partial x} \frac{\partial^2 \psi}{\partial x^2} - 
    \psi^* \frac{\partial^3 \psi}{\partial x^3} \right)  \right].
    \label{dim-less-total-j}
\end{eqnarray}
Here
$$
	u=\frac{8e^2\Gamma\psi_0^2}{\hbar^2\sigma_n}\frac{|K|}{|\alpha|}
$$
is a positive parameter, which can also be written as $u=\xi^2/l_E^2$ using the expression (\ref{l_E-def}) for the electric field penetration depth.

It should be noted that Eqs. (\ref{TDGL-eq-modified}) and (\ref{total-j-modified}) have not been derived from a microscopic theory. We expect that, similarly to the usual case, see Ref. \onlinecite{Kopnin-book},
the FFLO version of the TDGL formalism can be rigorously justified only under some restrictive assumptions, for certain values of the parameter $u$. Notwithstanding these reservations, we will follow a considerable 
precedent in the BCS case and use Eqs. (\ref{dim-less-TDGL-eq}) and (\ref{dim-less-total-j}) as a basis of a phenomenological theory of nonequilibrium FFLO superconductors.

\subsection{Stability of the normal state}

As an illustration of the TDGL formalism, in this subsection we investigate the local stability of a current-carrying normal state in an infinite wire below the critical temperature against the formation 
of the FF superconducting state. An analogous issue in the non-FFLO case was addressed in Refs. \onlinecite{Gor70}, \onlinecite{Kul71}, and \onlinecite{IKM80}. 

The time evolution of a small-magnitude nucleus of the FF state in a normal wire with $I\neq 0$ is determined by the linearized version of Eqs. (\ref{dim-less-TDGL-eq}) and (\ref{dim-less-total-j}), 
which has the following form:
\begin{equation}
	u\bigg(\frac{\partial \psi}{\partial t} + i I x \psi \bigg) = -\zeta \frac{\partial^4 \psi}{\partial x^4} - \frac{\partial^2 \psi}{\partial x^2} + \psi.
    \label{linear-mod-TDGL}
\end{equation}
In terms of the Fourier transform, $\psi(x,t) = \int_{-\infty}^{\infty}(dk/2\pi)e^{ikx}\Phi(k,t)$, we have
\begin{equation}
	\frac{\partial \Phi}{\partial t} - I \frac{\partial \Phi}{\partial k}  = Q(k)\Phi,
	\label{FT-TDGL}
\end{equation}
where 
$$
  Q(k) = \frac{1+ k^2-\zeta k^4}{u}.
$$
Note that in the usual BCS case the order parameter satisfies the same equation (\ref{FT-TDGL}), but with $Q(k) = (1-k^2)/u$. 

Solving the initial value problem for Eq. (\ref{FT-TDGL}) by the method of characteristics, we obtain:
\begin{equation}
	\Phi(k,t) = \Phi_0(k+It) e^{-S(k,t)},
	\label{Phi-general-solution}
\end{equation}
where $\Phi_0(k)=\Phi(k,0)$ and
$$
  S(k,t)=-\frac{1}{I}\int_{k}^{k+It} dk'Q(k')=\frac{\zeta}{5uI}\left[(k+It)^5-k^5\right]-\frac{1}{3uI}\left[(k+It)^3-k^3\right]-\frac{t}{u}. 
$$
Therefore, a small initial fluctuation of the order parameter evolves into
\begin{equation}
	\psi(x,t) = \int_{-\infty}^{\infty} dx'\, {\cal K}(x,x';t)\psi(x',0),
	\label{psi-general-solution}
\end{equation}
where the time evolution kernel is given by
\begin{equation}
\label{evolution-kernel}
      {\cal K}(x,x';t)=\exp(-iIx't) \int_{-\infty}^{\infty} \frac{dk}{2\pi}\, e^{ik(x-x')}e^{-S(k,t)}.
\end{equation}
Below we focus on the fate of the superconducting nucleus at $t\to\infty$.

The asymptotics of the momentum integral in Eq. (\ref{evolution-kernel}) can be evaluated using Laplace's method. At $t\to\infty$, the exponent in the integral has a minimum at $k=k_0=-It/2$,
in the vicinity of which
$$
  S(k,t)\simeq \frac{\zeta I^4t^5}{80u}+\frac{\zeta I^2t^3}{2u}(k-k_0)^2.
$$
Therefore,
\begin{equation}
\label{K-asymptotic}
  {\cal K}(x,x';t\to\infty)\simeq \sqrt{\frac{u}{2\pi \zeta I^2 t^3}} \exp\left(-\frac{\zeta I^4 t^5}{80u}\right) \exp\left[-\frac{u(x-x')^2}{2\zeta I^2t^3}-i\frac{It}{2}(x+x')\right].
\end{equation}
Due to the presence in this last expression of the rapidly decreasing exponential factor, any small fluctuation of the FF superconducting phase eventually dissipates, so that the current-carrying normal 
state is locally stable at all values of $I$, even below the critical temperature. Physically, a small-magnitude superconducting nucleus cannot screen the electric field, the latter causing acceleration and 
destruction of any incipient Cooper pairs.\cite{Gor70,Kul71} Superconducting state can possibly develop through the formation of a superconducting fluctuation of a finite magnitude (the critical nucleus) below a certain threshold
current, similarly to the non-FFLO case.\cite{KB77,Watts81,DBDF98}

\section{Conclusions}
\label{sec: Conclusions}

The FFLO superconductors, in which the Cooper pairs have a finite wavevector, can be described phenomenologically by a modified GL gradient energy. This modification profoundly 
changes the way a superconducting wire responds to an external current source and leads to a number of interesting effects. In this paper we have focused on the properties of the FF superconducting states, characterized by 
a single-plane-wave modulation of the order parameter. 

In contrast to the usual BCS case, there exist two distinct stable branches of the current-carrying FF superconducting states, bounded by two 
values of the critical current. Thermodynamics of a current-biased superconducting wire, in particular, the relative stability of the different FF states, is analyzed using the Gibbs free energy ${\cal G}$, instead of the Helmholtz
free energy ${\cal F}$. An external current can cause the wire to switch between the two branches through a first-order phase transition, which can be used experimentally as an evidence of the FF state.     
  
The twofold degeneracy of the superconducting ground state at zero external current leads to the possibility of novel topological defects -- the FF domain walls. At a nonzero current, the domain wall motion
may be detected by measuring a voltage across the wire. We have also shown that a ring made out of an FF superconductor can generate a spontaneous supercurrent in the absence of any external Aharonov-Bohm magnetic flux.

We have studied some nonequilibrium properties, using the phenomenological TDGL formalism with the modified gradient terms. We have shown that (i) a constant electric field penetrates 
an FF superconductor in exactly the same way as in the usual BCS case, and (ii) the current-carrying normal state is locally stable against the formation of the FF superconducting state at all currents.

\acknowledgments
We thank A. I. Buzdin for useful correspondence. 
This work was supported by a Discovery Grant (K. V. S.) and by a USR Award (B. P. T.), both from the Natural Sciences and Engineering Research Council of Canada.

\appendix

\section{FF state vs LO state}
\label{app: FF vs LO}

The general solution of Eq. (\ref{linear-GL}) is given by a superposition of two plane waves:
\begin{equation}
	\psi(x) = \Delta_1e^{iq_{0}x} + \Delta_2e^{-iq_{0}x},
	\label{FFLO-gen-solution}
\end{equation}
where $q_0$ is defined in Eq. (\ref{FFLO-wavevector}). Substitution of the above expression into the free energy density (\ref{modified-GL-energy-density}) yields
$$
	\frac{\cal F}{L} = a(T-T_c)(|\Delta_1|^2 + |\Delta_2|^2)+\frac{\beta}{2}\left[ (|\Delta_1|^2 + |\Delta_2|^2)^2 + 2|\Delta_1|^2 |\Delta_2|^2 \right],
$$
where all integrals of the oscillating terms are assumed to vanish. Parameterizing the plane wave amplitudes as $\Delta_1 = \Delta \sin \Theta$, $\Delta_2 = \Delta e^{i \Phi} \cos \Theta$,
we obtain:
$$
	\frac{\cal F}{L} = a(T-T_c)\Delta^2 + \frac{\beta}{2} \left(1 + \frac{1}{2}\sin^2 2\Theta\right)\Delta^4.
$$
At $\beta>0$, the minimum of the free energy is achieved when $\Theta$ is an integer multiple of $\pi/2$. Therefore, the most energetically favorable superconducting state corresponds to either $\Delta_1$ or 
$\Delta_2$ in Eq. (\ref{FFLO-gen-solution}) vanishing, \textit{i.e.}, to the FF state.

\clearpage

\clearpage

\clearpage

\clearpage

\clearpage

\clearpage

\clearpage

\clearpage

\end{document}